\documentstyle{article}
\makeatletter
\@addtoreset{equation}{section}
\def\@eqnnum{(\arabic{section}.\arabic{equation})}
\makeatother
\begin{document}
\begin{center}
\LARGE
\bf
One Proposal about the nature during \\
our time and space \\
\vspace{0.3em}
\large
Koji Ichidayama\\
\vspace{0.3em}
\it
Okayama716-0044,Japan\\
E-mail:ichikoji@lime.ocn.ne.jp\\
\vspace{0.3em}
1999.12.31\\
\end{center}
\large
\bf
Abstract. \
\rm
It made that the new symmetric property, the binary law, existed newly 
in our time and space at the thing except the symmetric property of the 
principle of general relativity which is already known in this paper 
clear.\\
The introduction of this symmetric property will have made dealing with 
the position tensor of us handy as much as the surprise.\\
\large
\section{Introduction}
\large
\ The new symmetric property during the time and space of us who 
propose by this paper is equal to the symmetric property of the already 
known principle of general relativity.\ It is a symmetric property 
with big scale as to be.
\ This symmetric property is caused by the mathematical law in the 
purity than, the physical law, the symmetric property about the dealing 
with the summation of the tensor.\ In the following chapter, it 
considers this thing and it proves a symmetric property about the 
dealing with the summation of the tensor first and it makes a new 
symmetric property during our time and space clear using this next.\\
\large
\section{One Proposal about the nature during our time and space}
\large
\ First, it considers about the symmetric property about the dealing 
with the summation of the position tensor.\\
When showing the ingredient $x^1, x^2, x^3, x^4$ of the position vector 
during our time and space in $x^\mu$ \cite{Dirac}, it decides to show 
the summation of these ingredients as follows.\\
\begin{equation}
X^\mu=\sum_{\mu=1}^{4}x^\mu
\end{equation}\\
Incidentally, because each of the ingredients are independence, they may 
use either of $X^\mu, x^\mu$ when showing a position vector but suppose 
that it uses $X^\mu$ in this paper.\\
The (2.1) formula can be shown as follows when differentiating (2.1) 
formula first here with ingredient $x^1$ and doing the operation to 
integrate again next about $x^1$ in $x^4$.\\
\begin{equation}
X^{\mu}=x^1,X^{\mu}=x^2,X^{\mu}=x^3,X^{\mu}=x^4
\end{equation}\\
Incidentally, in the (2.2) formula, it used the nature that the 
ingredient $x^1, x^2, x^3, x^4$ of the vector are independence each 
other.\\
By the way, it finds that it is possible to treat more in the same way 
as the scalar by position vector $X^\mu$ than the (2.2) formula 
here.\ Even if it replaces a scalar in the formula with the vector 
when the scalar of the vector differentiates, the logical contradiction 
doesn't occur to the equation and if using this thing, it finds the fact 
that it is possible to express in the following formula.\\
\begin{equation}
dX^{\mu}=\frac{dX^{\mu}}{dX^{\nu}}dX^{\nu}
\end{equation}\\
This formula is the differential calculus relational expression among 
two position vectors.\\
If using this style, two position tensors $X^{\mu}, X^{\lambda}$ can be 
expressed by the following relational expression.\\
\begin{equation}
X^{\mu}=\frac{dX^{\mu}}{dX^{\nu}}X^{\nu}
\end{equation}\\
\begin{equation}
X^{\lambda}=\frac{dX^{\lambda}}{dX^{\tau}}X^{\tau}
\end{equation}\\
Therefore, the relational expression to this summation of two position 
tensors is from (2.4), the (2.5) formula.\\
\begin{equation}
X^{\mu}+X^{\lambda}=\frac{dX^{\mu}}{dX^{\nu}}X^{\nu}+\frac{dX^{\lambda}}{dX^{\tau}}X^{\tau}=\frac{dX^{\mu}}{dX^{\nu}}X^{\nu}+\frac{dX^{\lambda}}{dX^{\nu}}X^{\nu}
\end{equation}\\
From the inner product of the vector and the distribution law about the 
outer product. \cite{Spiegel}\\
\begin{equation}
\left[\frac{d(X^{\mu}+X^{\lambda})}{dX^{\nu}}\right]X^{\nu}=\left[\frac
{dX^{\mu}}{dX^{\nu}}+\frac{dX^{\lambda}}{dX^{\nu}}\right]X^{\nu}=\frac{dX^{\mu}}{dX^{\nu}}X^{\nu}+\frac{dX^{\lambda}}{dX^{\nu}}X^{\nu}
\end{equation}\\
Incidentally, the vector in this place is a tensor.\\
If using this, the (2.6) formula is \\
\begin{equation}
X^{\mu}+X^{\lambda}=\frac{d(X^{\mu}+X^{\lambda})}{dX^{\nu}}X^{\nu}
\end{equation}\\
If comparing with the (2.4) formula, it finds that it is possible to 
treat the summation $X^{\mu}+X^{\lambda}$ of the position tensor in the 
same of one tensor.\\
From in the same way on the other hand, (2.4), the (2.5) formula of the 
relational expression to the product of two position tensors\\
\begin{equation}
X^{\mu}X^{\lambda}=\frac{dX^{\mu}}{dX^{\nu}}\frac{dX^{\lambda}}{dX^{\tau}}X^{\nu}X^{\tau}
\end{equation}\\
If comparing with the (2.4) formula, the product $X^{\mu}X^{\lambda}$ of 
the position tensor is different from the case with summation and it 
isn't possible to be treated in the same of one tensor by 
it.\vspace{0.3em}\\
Here, it decides to examine (2.8) formula in detail.\ When first, 
transform (2.8) formula and differentiate both sides in $X^\nu$\\
Incidentally, this treatment in the case is the same with the scalar.\\
\begin{equation}
\frac{dX^{\nu}}{dX^{\nu}}=\frac{d^2X^{\nu}}{d(X^{\mu}+X^{\lambda})dX^{\nu}}(X^{\mu}+X^{\lambda})+\frac{dX^{\nu}}{d(X^{\mu}+X^{\lambda})}\frac{d(
X^{\mu}+X^{\lambda})}{dX^{\nu}}
\end{equation}\\
Because it is a clause of 1st of the right side, 
$\frac{dS}{d(X^{\mu}+X^{\lambda})}(X^{\mu}+X^{\lambda})=0$ and 
moreover both sides are scalar quantity together and can change left 
side suffix $\nu$ into $\mu$\\
\begin{equation}
\frac{dX^{\mu}}{dX^{\mu}}=\frac{d(X^{\mu}+X^{\lambda})}{d(X^{\mu}+X^{\lambda})}
\end{equation}\\
It finds that there is not hinderance even if ${\mu}\neq{\lambda}$ meets 
actually here because the formula of either of ${\mu}={\lambda}, 
{\mu}\neq{\lambda}$ of (2.11) stands up and treats as 
${\mu}={\lambda}$ in case of treatment.\\
In other words, actually, the summation of the position tensor can be 
simplier treated.\\
Incidentally, it is possible to say that the symmetric property of 
${\mu}={\lambda}$ exists to the treatment of the summation 
$X^{\mu}+X^{\lambda}$ of the position tensor about this thing, 
too.\vspace{0.3em}\\
\hspace*{0.5mm} By the way, it considers about how a summation with the position 
tensor which is a place of the substantial adaptation of the symmetric 
property in case of this treatment next is expressed by the equation.\\
\ The summation of the position tensor which shows a position on the time 
and space which is different here in the future decides to show the 
suffix of the tensor as the uppercase of the English letter.\\
Also, it decides to show the summation of all position tensors which 
compose our whole time and space specifically in $X^I$.\vspace{0.3em}\\
First, if limiting relation to the relation of the position tensors, the 
relation that it is possible that it is possible to have some position 
tensor $X^{\mu}$ is limited to the relation among all position tensors 
except $X^{\mu}$.\\
This thing can be shown in $X^{\mu}=f(\overline{X^{\mu}})$ if deciding 
to represent $X^{\mu}$ and the position tensor which is independence as 
being $\overline{X^{\mu}}$.\\
Because it isn't clear here about the concrete form of 
$f(\overline{X^{\mu}})$, the definite relational expression about 
$X^{\mu}$ can not be gotten.\\
However, because the case where the relational expression can be fixed 
as prima facie only by the only mathematical request is only one, it 
makes clear about this case.\vspace{0.3em}\\
Because $\overline{X^I}$ doesn't exist, it becomes $\overline{X^I}=0$ 
to only $X^I$.\\
$X^I=f(0)$ here, therefore, because $X^I$ forms no kind of connection, 
the following formula can be set by it from $X^I=f(\overline{X^I})$.\\
\begin{equation}
X^I=0
\end{equation}\\
Here, such nature is only pure mathematical request like the (2.12) 
formula because it is not and can not form an equation to the summation 
of the position tensor except $X^I$.\ Therefore, the place of the 
substantial adaptation of the symmetric property in case of previously 
shown treatment is only the relational expression which was concerned 
with $X^I$.\vspace{0.3em}\\ 
\hspace*{0.5mm} Next, it considers about making a symmetric property in case of 
treatment which was previously shown to the (2.12) formula be adaptable 
actually.\\
First, it takes the summation about the position tensor to be treating 
by the (2.8) formula and it decides to show this in $X^J$.\\
\begin{equation}
X^{\mu}+X^{\lambda}+X^{\nu}=X^N+X^{\nu}=X^J
\end{equation}\\
The symmetric property in case of treatment which was previously shown 
here can be handled as follows if adaptable.\\
\begin{equation}
X^J=X^\mu+X^\nu
\end{equation}\\
This thing shows that it is possible to treat by making all tensors 
except $X^{\nu}$ the same in $X^{\mu}$ in case of treatment of 
$X^J$.\vspace{0.3em}\\
By the way, next, it thinks of $X^J$ and it thinks of extending in 
$X^I$.\\
First the relation between $X^J$ and $X^I$\\
\begin{equation}
X^I=X^J+X^M
\end{equation}\\
If therefore, substitute (2.13) formula\\
\begin{equation}
X^I=X^N+X^M+X^{\nu}=X^K+X^{\nu}
\end{equation}\\
Here, because it thinks that $X^N$ changed to $X^K$ only in case of 
$X^I$ and that the discussion to the (2.11) formula stands up in case of 
$X^K$, the symmetric property in case of previously shown treatment can 
be more handled as follows if adaptable than (2.13) formula.\\
\begin{equation}
X^I=X^\mu+X^\nu
\end{equation}\\
Therefore, $X^I$ can be $X^J$ in the same way treated and in case of 
treatment of $X^I$, all tensors except $X^{\nu}$ can be treated by 
making them the same in $X^{\mu}$.\vspace{0.3em}\\
By the dealing which shows (2.12) formula therefore, in (2.17)\\
\begin{equation}
X^\mu+X^\nu=0
\end{equation}\\
This thing shows that it is possible to treat by making all tensors 
except $X^{\nu}$ the same in $X^{\mu}$ in all position tensors which 
compose our whole time and space.\\
The introduction of this symmetric property will have made dealing with 
the position tensor of us handy as much as the surprise.\vspace{0.3em}\\

A symmetric property in case of treatment of these us in the time and 
space is called a binary law in the future and the tensor to be 
considering this rule is called BINOR.\\
\Large
\section{Discussion}
\large
\ It made that the new symmetric property, the binary law, existed 
newly in our time and space at the thing except the symmetric property 
of the principle of general relativity which is already known in this 
paper clear.\ The introduction of this symmetric property will have 
made dealing with the position tensor of us handy as much as the 
surprise.\\
Moreover, it decided to call the tensor which satisfies this binary law 
this back BINOR.\vspace{0.3em}\\
\hspace*{0.5mm} Incidentally, it plans to make the nature which is peculiar to BINOR 
clear with the continuing separate sheet of paper to this back this.\\

\begin{thebibliography}{99}
\bibitem{Dirac} P.A.M.Dirac,General Theory of Relativity (1975 by John 
Wiley \& Sons,Inc.)
\bibitem{Spiegel} Murray.R.Spiegel,Theory and Problems of Vector 
Analysis (1974 by McGraw-Hill,Inc.)
\end{thebibliography}
\end{document}